\documentclass[aps,prb,twocolumn,groupedaddress,showpacs]{revtex4}
\usepackage{longtable}
\usepackage{graphicx}% Include figure files
\usepackage{dcolumn}% Align table columns on decimal point
\usepackage{bm}% bold math
\newcommand{\LNS}{field induced nodal energy scale}
\newcommand{\ASJ}{Andreev - Saint-James}
\begin{document}

%Title of paper
\title{Field Induced Nodal Order Parameter in the Tunneling Spectrum of YBa$_2$Cu$_3$O$_{7-x}$ Superconductor}

% repeat the \author .. \affiliation  etc. as needed
% \email, \thanks, \homepage, \altaffiliation all apply to the current
% author. Explanatory text should go in the []'s, actual e-mail
% address or url should go in the {}'s for \email and \homepage.
% Please use the appropriate macro foreach each type of information

% \affiliation command applies to all authors since the last
% \affiliation command. The \affiliation command should follow the
% other information
% \affiliation can be followed by \email, \homepage, \thanks as well.
\author{G. Leibovitch} \email{guyguy@post.tau.ac.il}
\author{R. Beck} \thanks{Both first authors contributed equally to the paper} \thanks{Current address: University of California, Santa Barbara}  
\author{Y. Dagan}
\author{S. Hacohen}
\author{G. Deutscher}
\affiliation{School of Physics and Astronomy, Raymond and Beverly
Sackler Faculty of Exact Sciences, Tel-Aviv University, Tel Aviv,
69978, Israel}

%Collaboration name if desired (requires use of superscriptaddress
%option in \documentclass). \noaffiliation is required (may also be
%used with the \author command).
%\collaboration can be followed by \email, \homepage, \thanks as well.
%\collaboration{}
%\noaffiliation

\date{\today}

\begin{abstract}
We report planar tunneling measurements on thin films of
YBa$_2$Cu$_3$O$_{7-x}$ at various doping levels under magnetic
fields. By choosing a special setup configuration, we have probed a field induced 
energy scale that dominates in the vicinity of a node of the $d$-wave
superconducting order parameter. We found a high doping sensitivity for this energy scale.
At Optimum doping this energy scale is in agreement
with an induced $id_{xy}$ order parameter. We found that it can be followed down to low fields
at optimum doping, but not away from it.
\end{abstract}

% insert suggested PACS numbers in braces on next line
\pacs{74.72.Bk, 74.50.+r}
% insert suggested keywords - APS authors don't need to do this
%\keywords{}

%\maketitle must follow title, authors, abstract, \pacs, and \keywords
\maketitle
\section {Introduction}
It is by now well established that cuprate superconductors have an
order-parameter with a dominant $d$-wave symmetry, characterized
by node-lines along the [110] and equivalent directions\cite{tsuei:969}.
Nodal quasi-particles become then the
dominant low energy excitations \cite{Millis}. Interesting
phenomena have been predicted to occur when the $d$-wave
superconductor is subjected to a magnetic field perpendicular to
the superconducting planes: an energy gap should develop at the nodes,
that increases with the square root of the applied field.
This has been explained by Laughlin \cite{Laughlin} who assumes an additional
imaginary $id_{xy}$ component which increases with the magnetic field. Discrete
Landau energy levels were also predicted to develop in the nodal regions by Gor'kov-Schrieffer
\cite{gorkov} and Anderson \cite{Anderson:1998}.
However, the observation of nodal finite energy levels has encountered theoretical and experimental difficulties.
\par
It has been argued, that in the mixed state, superfluid screening
currents result in a Doppler-shift of the Landau levels larger than the level spacing,
rendering their observation impossible
\cite{gorkov,Janko,melnikov,Franz}. This Doppler shift will also
obscure a possible $id_{xy}$ component.\cite{Aubin}. Tunneling
experiments performed along a nodal direction, which should in
principle be an ideal method to probe nodal states, are dominated by low energy Andreev - Saint James (ASJ) surface states
due to the $d$-wave symmetry, resulting in a characteristic Zero Bias Conductance Peak (ZBCP).
The degeneracy of the \ASJ~ states is lifted by
screening currents splitting the ZBCP, an effect that can be confused with that of
nodal finite energy levels. Spontaneous time reversal
symmetry breaking effects are also sometimes observed
\cite{Lesueur:1992,Covington,Krupke,deutscherRMP,DaganPRL}.
In addition, it is not trivial to distinguish between the predictions of the Landau States
and of the minority order parameter theories. This is because,
the energy of the first Landau level is equal to the amplitude
of the $id_{xy}$ component predicted by Laughlin.
\par
To address these difficulties we have used field cycles that enable us to
distinguish between the Doppler shift and other possible spectral
contributions. We have concluded that data taken in decreasing fields is essentially free of Doppler shift effects and can be used to identify finite energy nodal states.  We confirmed that this energy follows the predicted square root of field behavior at optimum doping. By extending our measurements to 20 mK, we were able to follow the evolution of these states down to fields of the order of a few 1000 Gauss, where the Landau level interpretation is excluded due to the long length of the corresponding trajectories. Finally we report that the doping level had a strong influence on the splitting behavior of the ZBCP.
The square root behavior is not obeyed at low fields when deviating from optimum doping.
Such deviations were reported before \cite{DaganPRL,FSR},
however, the data was taken in increasing fields and was an admixture of the studied phenomena and Doppler shift.
In the new data presented and studied in this paper, the Doppler shift contribution
is essentially eliminated, and the doping effect appears far more clearly.
In Summary, our results favor the existence of additional $id_{xy}$ order parameter component
predicted by Laughlin rather than that of the formation of nodal energy levels predicted by Gor'kov-Schrieffer
\cite{gorkov} and Anderson \cite{Anderson:1998}.
\par
This paper will be organized as follows: we begin with a
theoretical background of the Doppler-shift effect.
In Sec. \ref{experimentalSEC} we present our experimental
setup enabling us to distinguish between the two contributions.
Our tunneling results at optimum doping will be
shown together with the low temperatures measurements at 20mK (Sec. \ref{resultsSec}).
We then compare our results to theory in Sec. \ref{discussionSec}
and finish with our conclusions and findings.
\section{Theoretical background}
\begin{figure}
\includegraphics[width=0.5\hsize]{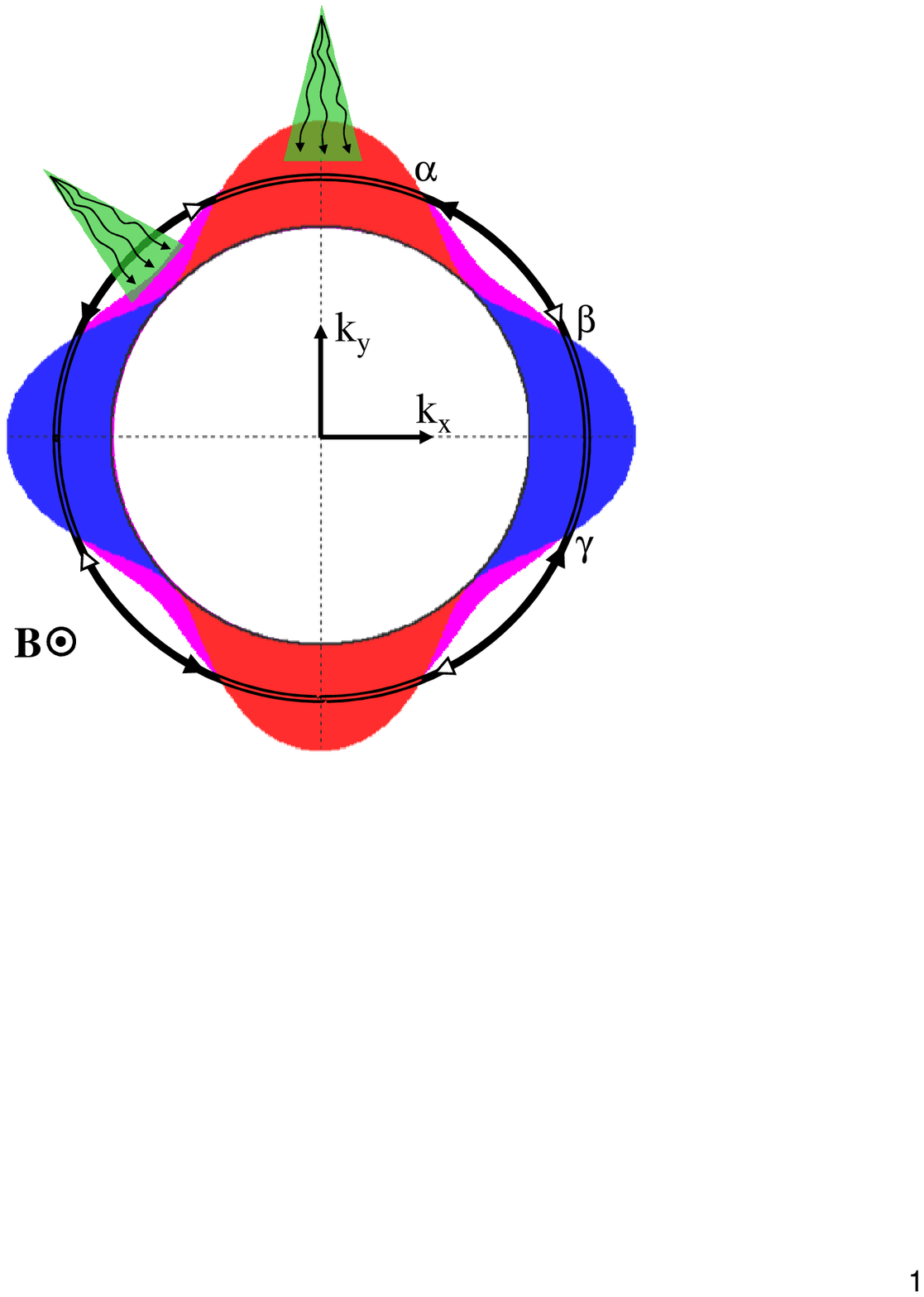}
\caption{(color online) Schematic illustration of the electronic
momentum space in a $d$-wave superconductor under an applied
magnetic field. The x and y axis are parallel to the [100] and [010] crystallographic directions perspectively.
For simplicity we assume: a cylindrical Fermi
surface, a  $d_{x^2-y^2}$-wave superconducting gap and nodes at $\pm
45^o$  from the principal axes, and a magnetic field, \textbf{B},
parallel to the z axis. The quasi-particles cycle of multiple
Andreev - Saint-James reflections forming the nodal energy level is marked by the solid line.
As an alternative theoretical description for the energy scale an
induced $id_{xy}$ order parameter (marked in purple) is predicted to
develop mainly in the vicinity of the nodes. The collimation of the
injected electrons in a planar tunneling configuration of the experiment is marked by
the green triangles for two different junction orientations. The
\LNS~can only be probed for tunneling along the node direction (left
triangle).\label{figscamatic}}
\end{figure}
\par
We wish to discuss the differences between two theoretical approaches
regarding the development of finite nodal energy states
under applied magnetic fields.
\par
In the first approach, by Laughlin \cite{Laughlin}, the free energy of a
$d$-wave superconductor subjected to a magnetic field
perpendicular to the superconducting planes can be minimized by
the inclusion of an additional $id_{xy}$ component to the main
$d_{x^2-y^2}$ component (illustrated in Fig. \ref{figscamatic} by the $id_{xy}$
order parameter marked in purple). The $id_{xy}$ component
breaks the symmetry in such a way that opposite currents will flow
on opposite faces of the sample creating a magnetic moment
parallel to the applied field. If the moment and the applied field are parallel
to each other the free energy will be minimized.
\par
In the second approach, following Anderson \cite{Anderson:1998},
we consider the motion of a quasi-particle in a nodal region (Fig. \ref{figscamatic}).
Under an applied magnetic field \textbf{B}, it acquires a velocity
component parallel to the Fermi surface. In the absence of a
superconducting order parameter, it will be in one of the Landau
levels, determined by the cyclotron frequency
$\omega_c=\frac{eB}{m^*c}$, where $e$ is the quasi-particle charge,
$c$ is the speed of light and $m^*$ is the effective electron mass. However, when
$\hbar\omega_c<\Delta$, where $\Delta$ is the amplitude of the
superconducting gap, the usual cyclotron motion is not possible.
Instead, as schematically described in Fig. \ref{figscamatic}, a
series of Andreev - Saint-James reflections in momentum space will
occur \cite{deutscherRMP}.
This process can only occur at certain energy levels
for which the total phase change during a Saint-James cycle is a
multiple of $2\pi$. These energy levels correspond values of
the current that flows around the Fermi surface.
\par
Interestingly, the amplitude of the minority component is the same
as the energy of the first Landau level \cite{gorkov,Anderson:1998}.
The two approaches lead to the same energy exactly:
\begin{equation}\label{eq1}
\varepsilon(B)=\pm 2\sqrt{\hbar\omega_c\Delta}
\end{equation}
Nonetheless, there are substantial differences between these two
approaches. While the Gorko'v-Schrieffer-Anderson theory assumes that the order
parameter is not altered by the magnetic field, its modification is a
key prediction in Laughlin's theory. The latter also predicts a
transition temperature above which the $d$-wave symmetry is
recovered. Finally Gor'kov-Schrieffer-Anderson predict a series of
energy levels while Laughlin only predicts a finite gap value.
\par
%Turning now to the experimental observation by tunneling spectroscopy
%of nodal states along a (110) direction, we first note that 
Tunneling along a nodal direction of a $d$-wave superconductor is
done through a surface where zero energy bound states are present
due to the interference of quasi-particles that undergo Andreev -
Saint-James reflections from lobes of the order parameter having
phases that differ by $\pi$ \cite{Hu,deutscherRMP}. These
bound-states should appear as a conductance peak at zero bias in
an in-plane tunneling spectrum \cite{TanakaPRL}. As shown by
Fogelstr\"om \emph{et al.}, this zero bias peak splits into two
spectral peaks due to a Doppler-shift from superfluid currents flowing
parallel to the surface \cite{FSR}. The peaks bias are
proportional to $v_s \cdot p_{_F}$, where $v_s$ is the superfluid
velocity and $p_{_F}$ the Fermi momentum of the probed states. For
example, when a magnetic field, $H$, is applied parallel to the surface,
Meissner currents Doppler-shift will produce spectral peaks which
are linear with $H$ up to a field of the order of the thermodynamical critical field 
where saturation is reached (about 1 Tesla in the case of YBa$_2$Cu$_3$O$_{7-x}$ (YBCO)).
\par
Since a nodal energy scale and a Doppler shift will both split the zero bias
conductance peak under an applied field, as has been observed \cite{Lesueur:1992,Covington,Krupke,DaganPRL}, one
must find a method to distinguish between both mechanisms and determine 
the difference in the predicted field dependences.
An obvious difference between them is that \LNS s are best
observed in the absence of superfluid currents, while a Doppler
shift effect exists only in their presence.
\par
A method, which we have already used \cite{Beck:2004}, consists in
performing magnetic field cycles. Meissner currents are quite
different in increasing and decreasing fields because the
Bean-Livingston barrier which can retard the penetration of
vortices through strong surface currents, up to a field of the
order of the thermodynamical critical field, is only effective
against flux penetration (increasing fields) and not against flux
exit (decreasing fields). As a result, strong surface Meissner
currents, on the scale of the London penetration depth, exist only
in increasing fields \cite{Bean:1968,Bussieres:1976,Clem}. 
Other types of current that can produce a Doppler shift are screening
currents around vortices \cite{graser:247001} and Bean's critical state currents \cite{Bean}.
%Another type of current that can produce a Doppler-shift is related to the
%presence of vortices below the surface \cite{graser:247001} and
%Bean's critical state currents \cite{Bean}.
The latter reversesign with field reversal, and in high fields, extends into the entire
thickness of the sample. They are typically weaker than the
Meissner currents in the Bean-Livingston regime.
\par
In planar tunneling experiments, electrons are injected across a
dielectric barrier into the superconductor. The transmission
probability decays exponentially with the increasing angle between the
electron's momentum and the normal to the interface resulting in a
collimated current. In a typical junction the momentum divergence
has a angle of 10-20 degrees known as the tunneling cone.
The width of this cone, which can vary slightly from one
junction to another, will influence the Doppler shift of the zero
energy surface bound states. however, it will not modify
the energy of nodal states nor that of an induced $id_{xy}$ order parameter component.
\par
In summary, a zero bias conductance peak is expected to appear in
a $d$-wave tunneling along the node direction and to split into
two spectral peaks when a magnetic field is applied perpendicular
to the CuO$_2$ planes due to a \LNS~ or via a Doppler-shift effect.
The two effects behave differently in a magnetic field. In the
next section, we show a way to minimize the Doppler-effect which allowed
us to probe the \LNS~ alone.

\section{\label{experimentalSEC}Experimental}
\begin{table*}
\caption{Samples characterization. As explained in the text, from
the resistivity temperature dependence measurement, R(T), we can
estimated the samples doping.  The zero field spectral peak bias
value is noted as $\delta_0$. T$_c$ is the temperature at zero
resistivity. The transition temperature width is determined by a Gaussian fit
to $\frac{dR}{dT}$ at the transition.}\label{tableSampels}
\begin{ruledtabular}
 \begin{tabular}{|c|c|c|c|c|c|c|c|c|}
  \hline
    Name & Figs. & Orientation & Thickness(\AA)& Doping regime & T$_c$(K) & T$_c$ width(K) & $\delta_0$(mV) & T(K)\\
   \hline
    S1 & 2,5,7 & (110) & 1600 & optimal & 88.1 & 1.4 & 0 & 0.3\\
    S2 & 3 & (110) & 1600 & optimal & 90 & 1 &  0 & 4\\
    S3 & 3 & (100) & 1000 & under & 84 & 1.5 & 0 & 4.2\\
    S4 & 4,7 & (110) & 1200 & optimal & 88.2 & 1.3 & 0 & 0.02\\
    S5 & 6 & (110) & 1200 & over & 87.3 & 1.5 & 1.4 & 4.2\\
    S6 & 7 & (110) & 600 & under & 87 & 1  & 0 & 4.2\\
    S7 & 7 & (110) & 1200 & over & 87.7 & 0.3 & 1.3 & 4.2\\
    S8 & 7 & (110) & 1200 & over & 85.7 & 1.1  & 1.6 & 4.2\\
    S9 & 7 & (110) & 1200 & over & 88.3 & 0.8  & 1.8 & 1.3\\
    S10 & 7 & (110) & 1200 & over & 86.7 & 0.7 & 1.8 & 1.3\\
    S11 & 7 & (110) & 1200 & over & 87.7 & 0.8 & 1.9 & 1.3\\
    S12 & 7 & (110) & 1200 & over & 88.3 & 0.6 & 1.5 & 1.3\\
    S13 & 7 & (110) & 1200 & over & 88.0 & 1 & 1.35 & 1.3\\
    S14 & 7 & (110) & 1200 & over & 87.9 & 0.3 & 2.25 & 1.3\\
  %  S15 & 7 & (110) & 1600 & optimal & 89 $\pm$ 2 & 0& 1.3 \\
     \hline
   \hline
\end{tabular}
\end{ruledtabular}
\end{table*}

Thin YBCO films were grown using DC off-axis sputtering
deposition. In order to minimize (103) oriented grains a buffer
layer of PrBa$_2$Cu$_3$O$_{7-x}$ was first deposited using RF
off-axis sputtering on top of the substrate \cite{Poelders}. We
used a SrTiO$_3$ and LaSrGaO$_4$ substrates for the (110) and
(100) oriented films respectively. These films have a well defined
[001] direction parallel to the surface of the film. $\theta
-2\theta$ x-ray diffraction patterns showed the relevant peaks for
the desired orientation. Scanning electron microscopy and atomic
force microscopy showed a well defined crystallographic growth and
surface roughness of a few nanometers. Resistivity measurements
showed the expected in-plane anisotropy.
In addition, the temperature dependence of the $ab$ plane
resistivity allowed us to estimate the doping level in our films.
It changes from a positive curvature for overdoped, linear with
temperature for optimally doped and negative curvature for
underdoped films \cite{DaganEPL2,PhysRevB.53.5848,PhysRevB.52.16246,castro:174511}.
I-V characteristics were measured using a current source and a digital voltmeter.
The tunneling conductance spectra were calculated by differentiating the I(V) curves.
Table \ref{tableSampels} shows the characterization values for the 15
representative junctions used in the figures of this paper.
\par
In our experiments the tunneling junction is created by placing an
indium electrode on top of the surface of freshly prepared thin
YBCO films. At the metal-superconductor
interface a thin insulating indium-oxide layer is then formed, which is stable
over weeks and many thermal cycles. We can verify the quality of
the tunneling junction by several methods. First, by lowering the
temperature below the indium critical temperature we can measure
the well-know indium tunneling spectrum which dominates the low
energy spectra. Also the indium spectrum disappears as we
increase the magnetic field or heat up the sample above the
indium's critical field and temperature, respectively. Second, we
ensure that the high bias conductance is insensitive to the magnetic 
field and temperature below the YBCO critical temperatureas, as expected for tunneling
spectroscopy.
\par
Our sample and the field configuration are favorable for several
reasons. First, the magnetic field is applied parallel to the surface (which avoids threading the tunnel junction
with vortices) and at the same time perpendicular to the CuO$_2$ planes, as required
for the observation of finite energy nodal states. Second, the undesired
Doppler-shift due to superfluid currents is minimized because in our
geometry $v_s \cdot p_{_F}$ is at a minimum since the dominant currents flow parallel to the interface. Third, by
comparing data taken in increasing and decreasing fields it is
possible to identify the effect of the Doppler-shift due to
strong surface Meissner currents that exist only in increasing fields
\cite{Bean:1968,Bussieres:1976,Clem}. Therefore, measurements in
decreasing fields are less affected by the Doppler-effect.
Also, the intensity of the Meissner currents themselves can be
minimized by using films whose thickness is smaller than the London
penetration depth \cite{Beck:2004,galgal}. Finally, we have performed experiments both 
on (110) and (100) orientated films. While for the [110] direction
we expect to probe the field induced energy scale, for the [100]
direction it should not be observed. We emphasize that in our method due to the planar
geometry of the junction and the resulting tunneling cone,
we probe a specific direction in $k$-space rather than the $\textbf{k}$ averaged local
density of states as probed by scanning tunneling microscopy.
\par
\begin{figure}
\includegraphics[width=1\hsize]{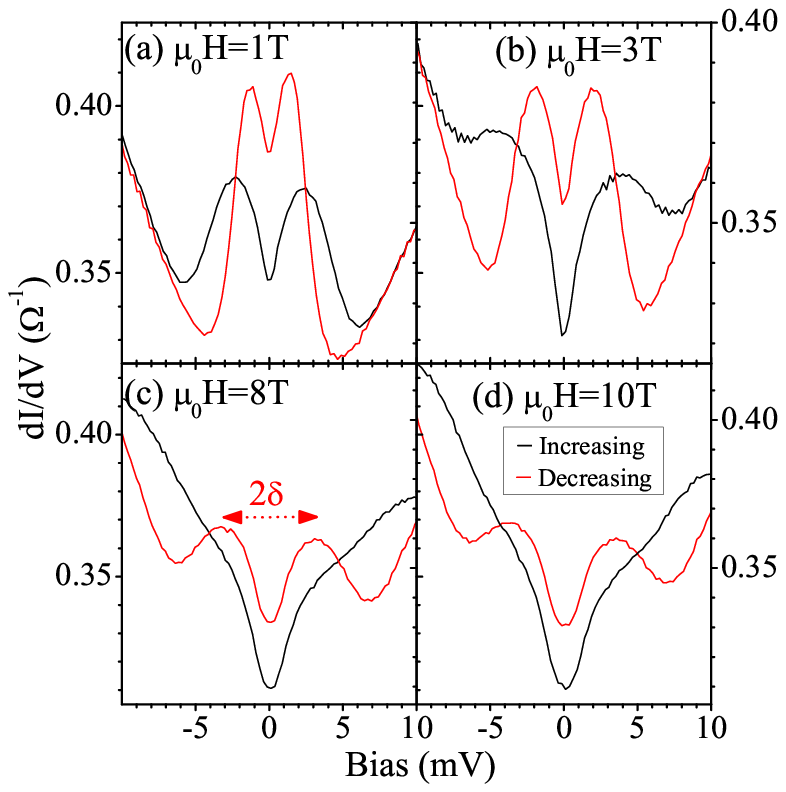}
\caption {(color online) Tunneling spectra, dI/dV(V), for
various magnetic fields (sample S1, at 0.3 K). Increasing from
zero magnetic field (black) and decreasing from 16 T (red), they
present different tunneling spectra in a given magnetic field. We
define $2\delta$ as the distance between the two maxima in the
spectrum. For a given magnetic field, $\delta (H)$ is always
larger in increasing fields than in decreasing one. This is due to
the Doppler-shift effect resulting in shifting and smearing the
\LNS~peaks to higher biases in increasing magnetic fields. In
fields higher than 4 T, the peaks can only be identified in
decreasing magnetic fields.\label{tunnelingSpectrum}}
\end{figure}

\par
\section{\label{resultsSec}Results}
\subsection{\label{optimallySubSec} Probing the \LNS}
In Fig. \ref{tunnelingSpectrum} we show tunneling spectra for an
optimally doped (110) oriented film as a function of magnetic field
applied perpendicular to the CuO$_2$ planes. We notice that the
spectral peak bias values, $\delta(H)$, are always larger in
increasing rather than in decreasing fields. In addition,
the peak seen in decreasing fields is well defined for all fields,
while that seen in increasing fields becomes very broad 
and is too broad to be identified (Fig. \ref{tunnelingSpectrum}), while in decreasing fields it remains
well defined up to more than 22 T (see for example Fig.
\ref{funnelfig}).
\par
\begin{figure}
\includegraphics[width=1\hsize]{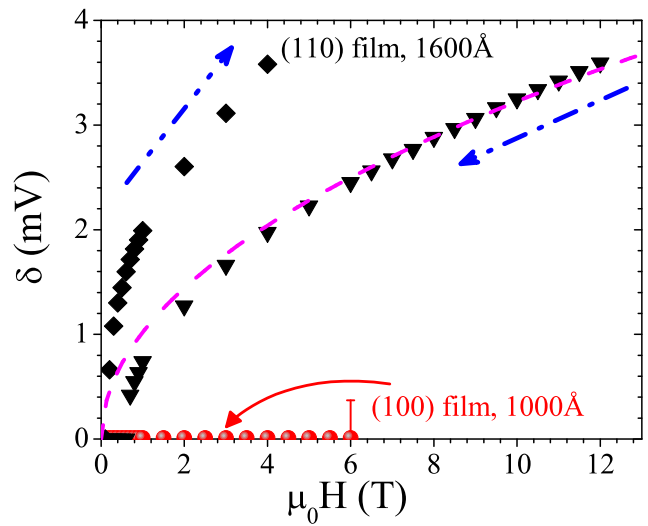}
\caption{(color online) Spectral peak bias value, $\delta$, as a
function of magnetic field at 4.2 K. Black triangles (diamonds)
represent the peak positions in decreasing (increasing) magnetic
fields for a (110) oriented film having thickness of 1600\AA
(sample S2). While a Doppler-shift in increasing fields prevents
observation of the \LNS, the measurements in decreasing fields are
free of the Doppler-shift effect and allow unambiguously
identification of the \LNS. Dashed purple line is a fit to
$\delta=AH^\frac{1}{2}$, with $A=1.02 \pm$ 0.05
meV/T$^\frac{1}{2}$ in excellent agreement with Eq.(1). The red
circles represent (100) oriented film with a thickness of 1000\AA
(sample S3). The \LNS~are not observed, as expected.
\label{sqrtHfig}}
\end{figure}
\begin{figure}
\includegraphics[width=1\hsize]{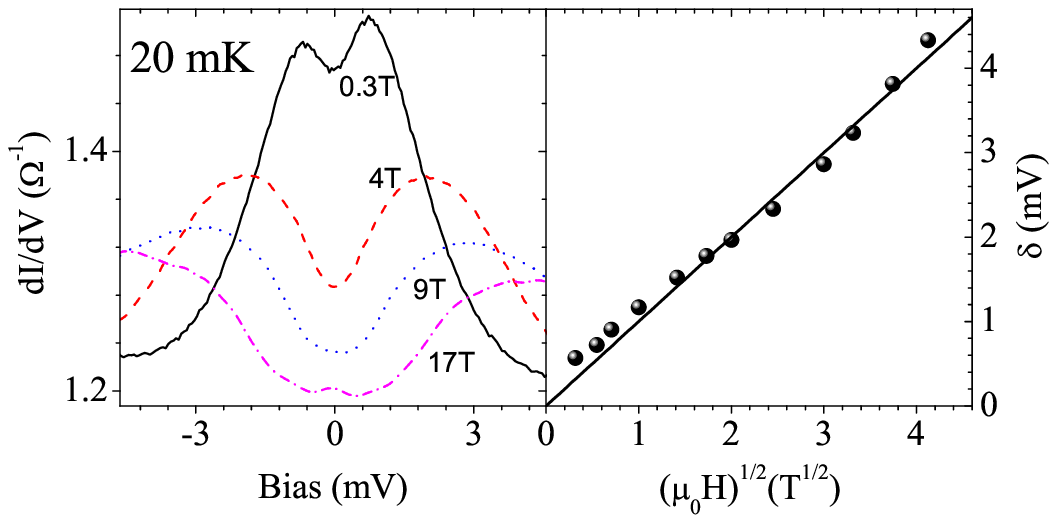}
\caption{(color online) (a) Tunneling spectra taken at 20mK in
decreasing magnetic fields (sample S4). Note the constant
conductance in the vicinity of zero bias supporting the existence
of an energy scale $\varepsilon = \pm\delta(H)$ (b) Spectral peak value in decreasing
fields versus square root of the applied magnetic field measured
at 20mK. the line is a fit to high fields having a slope of 1.01
meV/T$^\frac{1}{2}$.\label{fig20mK}}
\end{figure}

We demonstrate in Fig. \ref{sqrtHfig}, and \ref{fig20mK} that the spectral peak
bias value does not increase linearly at low fields,
and does not saturate at high field. This is in contradiction with
the Doppler shift theory of Zero Energy Surface Bound States.
A substantial difference between the behavior of (100) and (110) oriented
films is observed. While for thin (110) films, the spectral peaks are
clearly seen even at low fields, no such peaks can be detected in thin (100) films.
Since in (100) films, the only splitting mechanism is the Doppler shift effect,
the absence of the spectral peaks in such films is indicative of the insignificance
of that effect in films thinner than the London penetration depth.
Upon increasing the (100) oriented films thickness, the contribution of the Doppler
shift is also increased and the spectral peaks are recovered \cite{Krupke}.

The third qualitative evidence supporting the argument that we
probe a \LNS~ rather than a Doppler-shifted zero bias
conductance peak comes from the shape of the tunneling
conductance at zero bias. While the Doppler shifted ZBCP results in a V-shape
conductance \cite{FSR,graser:247001}, one expects a U-shaped conductance at
zero bias for the \LNS~ \cite{PhysRevB.64.214519}. Because it is difficult to distinguish
between these two shapes due to thermal population effects, one should perform
measurements at very low temperatures. In Fig.\ref{fig20mK} we show
tunneling spectra taken at 20mK under high magnetic fields. The
observed U-shaped conductance is in agreement with the
\LNS~scenario and contrasts the Doppler shift model.

We have therefore demonstrated that a Doppler shift due to Meissner
screening currents does not play an important role in our tunneling measurements of thin
films in decreasing fields.
However, the effect of a nearby vortex \cite{graser:247001}, or Bean's critical
state currents \cite{Bean} could still take place in a decreasing
magnetic field. These currents reverse their polarity
when decreasing the magnetic field from the maximum value reached.
If they dominate, the total current should be zero at some field. In this case, a zero bias
peak should reappear at a finite magnetic field. Such a behavior was never observed, ruling 
out that the spectral peaks in decreasing fields arise from the effect of a
nearby vortex or a strong Bean critical state currents (see for example a field cycle in Fig. \ref{polarity2Fig}).

\par
Finally we note that the peak position measured in decreasing fields is
extremely reproducible, for a variety of samples and junctions.
This is in contrast with the expected behavior of the Bean critical currents
and the Doppler shift effect that are strongly dependent on film thickness,
tunneling cone and surface barrier formation.

\par
\par

In contrast with the difficulties encountered in trying to explain
the data shown in Fig. \ref{tunnelingSpectrum} and Fig. \ref{sqrtHfig}
by a Doppler shift of zero-energy surface bound-states, a \LNS~ provides a reasonable explanation.
As shown in Fig. \ref{fig20mK}, the data taken in decreasing fields fits the
predicted square root dependence on the magnetic field. The field
hysteresis shown in Fig. \ref{tunnelingSpectrum} and Fig.
\ref{sqrtHfig} can be understood as due to a Doppler shift by
Meissner currents in increasing fields; the peaks are
shifted to higher bias and are broadened until they cannot be
identified anymore in very high fields.

To summarize, we have shown that for an optimally-doped sample, tunneling
measurements on (110) oriented film reveal an energy scale that can be
understood either as an additional complex order parameter
in the form of $id_{xy}$ or as the first Landau state in the
vicinity of the $d$-wave node.

\par

\subsection{\label{GSAvsL} Implausibility of nodal Landau levels}

After ruling out the Doppler shift effect, we now concentrate on distinguishing
between the nodal Landau states approach of Gor'kov-Schrieffer \cite{gorkov}
and Anderson \cite{Anderson:1998} and the nodal order parameter component
predicted by Laughlin \cite{Laughlin}. Although the first Landau level and
the nodal order parameter have identical field dependences, the following evidences favor the later.

\par
An important feature is observed in Fig. \ref{fig20mK}:
only two peaks at $\pm\delta$ are visible. This is incompatible
with the theories in refs. \cite{gorkov,Anderson:1998}, which
predict a series of energy levels that should manifest
themselves as peaks in the tunneling spectrum. These
peaks have never been observed. One could argue that the higher
energy levels are hidden due to scattering. The visibility of the
first peak at relatively high temperatures ($T>4.2K$) under small
magnetic fields ($H\simeq 0.3T$) suggests that the scattering processes
are not strong enough to obscure the higher peaks at 20mK and 9T,
yet, these peaks are not observed in
Fig. \ref{fig20mK}.

\par
Further evidence comes from estimating the trajectory length at the node area.
If we define $\theta$ as the angle from the anti-nodal direction at which the particles meet the
superconducting gap $\Delta$ and perform an \ASJ~ reflection, and $\pm\alpha$
as the trajectory angle measured from the node we get $\alpha=2(45-\theta)$.
Using Eq. \ref{eq1} we calculate $\alpha$ for small angles to be
$\alpha=\frac{\varepsilon}{\Delta} = 2\sqrt{\frac{\hbar e B}{mc\Delta}}$.
In the node area, the trajectory can be treated classically
using the cyclotron frequency $\omega_c$. In the node vicinity, the trajectory time
$t=\frac{\alpha}{\omega_c}$ gives the trajectory length $x$:

\begin{equation}\label{trajlength}
x = tV_f = 2\sqrt{\frac{\hbar mc}{eB\Delta}}V_f
\end{equation}

where $V_f$ is the Fermi velocity.
Using Eq. \ref{trajlength}, we calculate that for $B=1$ Tesla, $\Delta=20meV$ and $V_f=10^6\frac{m}{sec}$
the trajectory length is $8600 \AA$ which is several times larger than the film's thickness.
Scattering from the surface will not allow the formation of Landau states, unless the applied field is of the order of 100T.
But in fact the nodal scale is observed at low temperatures (20mK) down to a fraction of a Tesla (Fig \ref{fig20mK}).

\par
Two additional indications favoring Laughlin's theory over Gor'kov-Schrieffer-Anderson
will be discussed in the the following sections:
the effects of doping in section \ref{dopingSubSec} and a first order phase transition
of the \LNS~ measured by Elhalel \emph{et al.} \cite{galgal} in section \ref{discussionSec}.
\par
\subsection{\label{dopingSubSec} Effect of doping on \LNS}
%\subsubsection{\label{nonconcomitant} non concomitant}

\subsubsection {\label{subsectionUnder}Underdoped case}
\begin{figure}
\includegraphics[width=1\hsize]{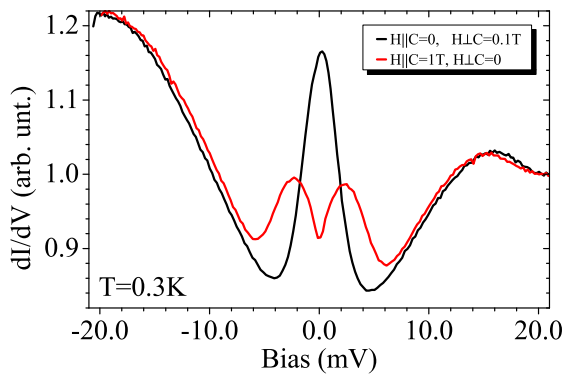}
\caption{Tunneling conductance for a (110) film taken at 0.3K
(sample S1). The black line measured in magnetic field of 0.1 T
applied parallel to the CuO$_2$ planes where the indium counter
electrode is in its normal state without field inducing spectral
peaks. The red curve is for 1 T in increasing fields applied
perpendicular to the CuO$_2$ planes, where \LNS~peaks are
observed.}\label{ZBCP03Fig}
\end{figure}

For underdoped samples at zero field only a single peak is
observed at zero bias (the ZBCP). We shall now demonstrate that
this is not due to the thermal smearing of two peaks at finite energy.
To do that we reduced the temperature to 0.3K while applying a
small magnetic field perpendicular to the \emph{c-axis} (and
parallel to the sample's surface). This field quenches
superconductivity in the indium counter electrode but has no
effect on the ZBCP as has been demonstrated by Krupke and
Deutscher \cite{Krupke}. The results are shown in Fig.
\ref{ZBCP03Fig}. The sample is slightly underdoped at $T_c=88.1 K$.
The black solid line shows a sharp zero bias peak without any
observable splitting. An upper limit for the bias of possible spontaneous peaks is
extracted from this measurement to be $2k_BT= 0.06meV$. We note
that even at 16~T, we do not observe any zero bias peak splitting
under this configuration, which ensures the sample's orientation.
By contrast, when the field is applied parallel to the sample's
\emph{c-axis}, the two spectral peaks at $\pm$2.2 meV are clearly
seen.

\begin{figure}
\includegraphics[width=0.8\hsize]{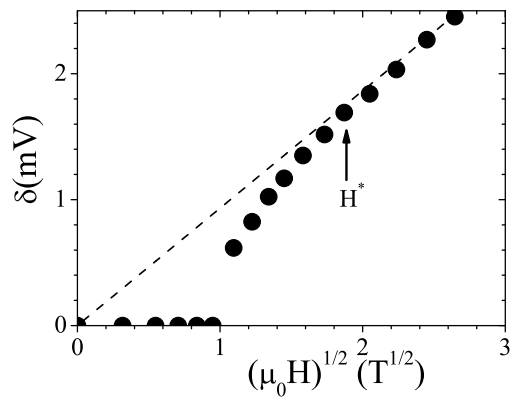}
\caption{Spectral peak value, $\delta$, as a function of the
square-root applied magnetic field (sample S6). The line is a
linear fit to the high magnetic field data. It has a slope of 0.9
meV/T$^\frac{1}{2}$.\label{underdopedFig}}
\end{figure}

In Fig. \ref{underdopedFig} we show a typical field dependence of the
nodal energy scale. At low fields up to 1 Tesla, a single peak is observed at zero
bias. Upon increasing the magnetic field a \LNS~ appears at
lower energies when compared to the case of optimum doping, until it reaches
a cross-over field (marked by $H^*$ in Fig. \ref{underdopedFig})
at which it recovers the optimally-doped $\sqrt{H}$ behavior. In unserdoped
samples there appears to be a well defined field below which the ZBCP 
does not split. This field increases rapidly with underdoping.

\subsubsection{\label{OverdopedSubSubSec}Overdoped case}

\begin{figure}
\includegraphics[width=1\hsize]{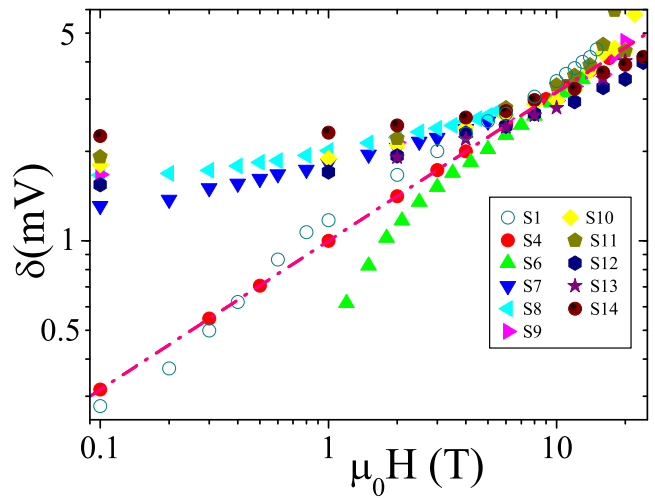}
\caption{(color online) Spectral peak positions, $\delta$, as a
function of magnetic field for various doping levels films at
log-log scale. All data shown here was taken in decreasing
magnetic field. In overdoped films the tunneling spectrum exhibits
peaks at zero magnetic field, where for underdoped ones peaks are
missing at low magnetic fields. All $\delta$ values coincide at high
magnetic fields and follow the dashed line having a slope of
1~meV/H$^\frac{1}{2}$. \label{funnelfig}}
\end{figure}

\par
All overdoped samples exhibit spontaneous zero field spectral
peaks. Dagan and Deutscher \cite{DaganPRL} have shown a
correlation between the spontaneous spectral peak bias values,
$\delta_0$, and the rate at which this bias increases with field.
They concluded that the spontaneous peaks are due to a
modification in the order-parameter symmetry near the surface in
the vicinity of optimum doping from pure $d-wave$ for underdoped
samples, to $d\pm id_{xy}$ or $d\pm is$ in overdoped ones.
\par
In this section we present a study of overdoped samples with
different oxygen doping levels in high \emph{decreasing} magnetic fields
starting from fields as high as 32.4 T. The results are summarized in Fig.
\ref{funnelfig}. For slightly overdoped films (see all data points
above the dashed line), we find zero field spectral peaks, with a minute
shift at low magnetic fields. The zero and low magnetic
field data are qualitatively in agrement with the measurements of
Dagan and Deutscher \cite{DaganPRL}. However, new behavior is
observed at high magnetic fields. At these high fields, all data
points collapse to a single line having a slope of
1~meV/H$^\frac{1}{2}$ (dashed line). This is the same slope as found
at high fields for optimally doped and underdoped samples.
%(dashed line). Next we show that other theories predicting zero
%field spectral peaks cannot explain our data.

\par
\begin{figure}
\includegraphics[width=1\hsize]{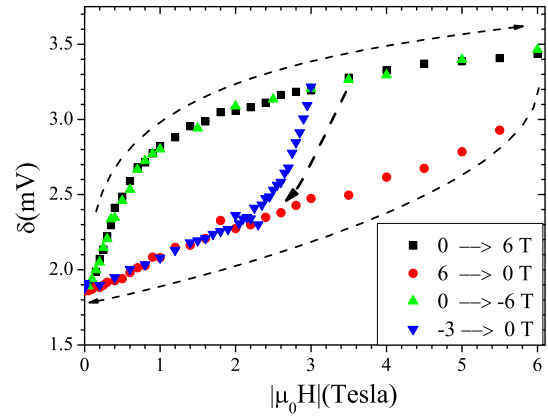}
\caption{(color online) Spectral peak position, $\delta$, as a
function of the absolute magnetic field in both polarities (sample
S5). We found no evidence that the polarity of the magnetic field
influences the peak position for a given
field.}\label{polarity2Fig}
\end{figure}

It has been suggested that finite energy peaks can result
from trapped vortices and their associated Doppler shifting
super-current at the surface \cite{graser:247001}. Another explanation 
could be a minority imaginary component of the superconducting
order-parameter \cite{FSR,fogelstrom2}.
%In such a case, when the
%order-parameter has an additional component $\pm is$ or $\pm
%id_{xy}$ component, it will spontaneously produce surface currents
%due to the phase gradient, resulting in spectral peaks in the
%tunneling spectrum \cite{TanakaPRL}.
Both cases break time
reversal symmetry, as the spontaneous currents flow in a specific
direction. Applying additional currents should then
result in either increasing or decreasing the net current,
assuming that the time reversal symmetry is broken
macroscopically.
%We stress that, in terms of energy, both plus and
%minus sign components are degenerate in zero external field.
However, we find no differences in the tunneling spectra for both
polarities. This is shown in Fig.~\ref{polarity2Fig}. When the
magnetic field is increased for a zero field cooled sample the
spontaneous peak value should reduce for about  one half of the
samples; or when the induced current is opposite to the
spontaneous one. However, in over 100 junctions measured in this
study we never observed that the spontaneous peak's bias
decreased with increasing magnetic field.

\par
Theoretical studies by Asano \emph{et al.}
\cite{asano:134501,asano:214509} and Kalenkov \emph{et al.}
\cite{kalenkov:184505}~show that surface impurities could also result
in spontaneous spectral peaks. These models predict that
scattering of impurities cause bound states that dominate
the low bias spectrum. However the \ASJ~ bound states
will still be present, and a three peaked structure at zero-bias
should be present. We did not observe such behavior in any of the
junctions presented here. Moreover, according to these models, the
``impurity'' spectral peaks bias value should be proportional to
the amount of impurities at the surface. In the case of a clean
junction, a zero bias peak should be present. However, when a
magnetic field is present, a splitting via an Aharonov-Bohm like
phase shift will occur \cite{asano:134501}. This means that at
high bias the zero field spectral peaks will be shifted by an
external magnetic field and for a clean interface,
no magnetic field shifted peaks should be observed.
However, an earlier study by Dagan and Deutscher
\cite{DaganPRL}(see also Fig.~\ref{funnelfig})
shows the opposite trend. At low magnetic fields,
junctions showing a spontaneous spectral peak shift to a lesser degree
than those showing a zero bias conductance peak. This rules out
impurities as a possible explanation for spontaneous peaks.

\section{\label{discussionSec}Discussion}
\par
As mentioned before both theories - Laughlin's \cite{Laughlin} $id_{xy}$ theory
and Landau-states by Gor'kov-Schrieffer - Anderson
\cite{gorkov,Anderson:1998} result in exactly the same field
dependence for the induced nodal scale. However, there are
two main differences between these approaches that can be checked
experimentally. First, Laughlin predicts a weak
first-order phase transition to the $id_{xy}$ state which is not
predicted by the Landau-state theorem. This phase transition was
in fact demonstrated recently by Elhalel \emph{et al.}
\cite{galgal}. Second, the Gor'kov-Schrieffer - Anderson theory predicts a series
of states while in Laughlin's theory, only one energy scale
appears. The second peak is not observed, even down to 20mK. Additionally, we showed that
the trajectories between two successive \ASJ~ reflections are much longer than
the films thickness. It is therefore unlikely that such states exist in the thin
films used in our measurements.

\par
Laughlin's theory however,has no doping dependence, which is a key feature observed in our measurements.
Following Laughlin, Deutscher \emph{et. al.} \cite{DeutscherMF} suggested a doping
dependence correction to the free energy in the form

\begin{equation}\label{GDfree}
F = a\delta^2 + b\delta^3 - c\delta B
\end{equation}

Here $b$ and $c$ are calculated by Laughlin. $a$ is a doping dependent term, $a = a_0(x-x_c)$,
where $a_0$ is a negative constant and $x_c$ is the optimal carrier concentration.

Using Eq. \ref{GDfree}, we calculate a minimum $F$ for $\delta\neq 0$ at zero field only
in the overdoped regime where $x>x_c$, while at higher fields the square root behaviour
of $\delta$ is recovered for both underdoped and overdoped regimes.
We can therefore conclude that our data is better described by the modified Laughlin's theory \cite{Laughlin,DeutscherMF}
with an additional order parameter component.

\par
%Next, we discuss the origin of the spontaneous spectral peak.
We have shown that time-reversal symmetry 
is not broken, macroscopically, even when spontaneous spectral
peaks appear in the tunneling measurements. An experiment, similar in
concept, was conducted by Tsuei \emph{et al.} where they measured
the spontaneous half flux-quantum vortex in a tri-crystal
experiment and found no difference between spontaneous vortices at
opposite polarities \cite{tsuei:187004}. The tri-crystal
experiment claimed to rule out a minority component to the
superconductor order-parameter.

\par
Because our measurements, as well as the tri-crystal ones, are
macroscopic, domain regions with alternating spontaneous
current directions can reconcile both experimental results.
The origin of such regions could be alternating minority
order-parameters having $\pm id_{xy}$ or $\pm is$ symmetries as
both plus and minus states are degenerate. In fact, such
configurations could be energetically favorable. In such a case, at
the domain wall region the spontaneous current should be zero. Our
technique has the advantage that it can probe the zero magnetic
field state, and the microscopic time-reversal symmetry breaking,
regardless of the spontaneous currents direction at the
microscopic scale.
\par
Since the order-parameter changes over a length scale set by the
superconductor coherence length, one should be able to find
nanometer scale regions (the domain wall region) where the
minority component order-parameter is zero, while, in other
regions (inside the domains), it should be finite. Only a microscopic
study, for example with a scanning tunneling microscope,
may be able to probe such domains. Furthermore, in the domain
wall region, the order-parameter symmetry should be purely
$d$-wave. Therefore, measurements aiming at detecting node-line
excitations, such as thermal conductivity, may be dominated by the domain wall regions.

\par
\section{\label{summarySec}Summary}
In conclusion, tunneling experiments revealed that the spectrum of
quasi-particle states in nodal regions of a $d$-wave superconductor
is profoundly modified by applying a magnetic field perpendicular to
the CuO$_2$ planes.
Doppler-shift of the \LNS~has been identified, and as expected,
is large enough to prevent observation of a \LNS, however it
has been minimized by choosing an appropriate geometry. We showed
that the zero field spectral peaks cannot be explained by either
inelastic scattering or trapped vortices at the surface, but rather by
a domain-like structure of minority order-parameter components with
alternating signs. We studied the interplay between the spontaneous
spectral peaks with the formation of the \LNS~and film doping.
Although the low energy states are in agreement with
theories based either on the explicit description of \ASJ~reflections
by the order parameter away from the nodes\cite{gorkov,Anderson:1998},
or on a free-energy expression that takes into account a gain in energy
due to the interaction between the applied field and the magnetic moment created by an
$id_{xy}$ component \cite{Laughlin}. But the absence of higher energy
level peaks and unreasonably long length of the trajectories that would be necessary to observe Landau levels are in contradiction with the former, while the doping dependance and the first order phase transition observed by Elhalel \emph{et al.}  \cite{galgal}, are in favor of the existence of a field induced $id_{xy}$ order parameter. 

% the absence of higher energy level peaks,
%the doping dependence and the first order phase transition observed by
%Elhalel \emph{et al.}  \cite{galgal} et .al, the doping dependence
%and the unreasonable length of the trajectories
%are all in favor of the existence of the $id_{xy}$ order parameter.

\begin{acknowledgments}
This work was supported by the Heinrich Herz-Minerva Center for
High Temperature Superconductivity and by the ISF. A portion of
this work was performed at the National High Magnetic Field
Laboratory, which is supported by NSF Cooperative Agreement No.
DMR-0084173, by the State of Florida, and by the DOE. We
acknowledge the assistance from Roman Mints, Alexander Gerber and
Enrique Gr\"{u}nbaum. G.D. wishes to acknowledge the hospitality
of Stanford University during the final preparation stage of this
work. Y.D. acknowledges the support from GIF. R.B. acknowledges
the hospitality of Richard Greene and the center for
superconductivity research at the University of Maryland, where
preliminary results for this work were obtained.
\end{acknowledgments}

\bibliographystyle{apsrev}
\bibliography{LandauStates}
\end{document}